\documentclass[a4paper,11pt]{article}
\usepackage{a4wide,amsmath,amssymb,bbm}
\usepackage[english]{babel}

\pdfoutput=1

\newcommand{\bea}{\begin{eqnarray}}
\newcommand{\eea}{\end{eqnarray}}

\newcommand{\nn}{\nonumber}

\newcommand{\ket}[1]{|{#1}\rangle}

\newcommand{\rf}[1]{(\ref{#1})}
\usepackage{graphicx}
\usepackage{feynmp}
\usepackage{color}

\def\unit{\protect{{1 \kern-.28em {\rm l}}}}

\DeclareGraphicsExtensions{.pdf}  

\makeatletter

\@addtoreset{equation}{section}\makeatother

\begin{document}

\newsavebox{\feynmanrules}
\sbox{\feynmanrules}{
\begin{fmffile}{diagrams} 


\fmfset{thin}{0.6pt}  
\fmfset{dash_len}{4pt}
\fmfset{dot_size}{1thick}
\fmfset{arrow_len}{6pt} 



\begin{fmfgraph*}(80,40)
\fmfkeep{schannel}
\fmfleft{i1,i2}
\fmfright{o1,o2}
\fmf{plain}{i1,v1}
\fmf{plain}{i2,v1}
\fmf{plain,left=0.5,tension=0.4}{v1,v2}
\fmf{plain,right=0.5,tension=0.4}{v1,v2}
\fmf{plain}{v2,o1}
\fmf{plain}{v2,o2}
\fmfdot{v1,v2}
\end{fmfgraph*}

\begin{fmfgraph*}(80,40)
\fmfkeep{tchannel}
\fmfleft{i1,i2}
\fmfright{o1,o2}
\fmf{plain}{i1,v1,o1}
\fmf{plain}{i2,v2,o2}
\fmf{plain,left=0.5,tension=0.4}{v1,v2}
\fmf{plain,right=0.5,tension=0.4}{v1,v2}
\fmfdot{v1,v2}
\end{fmfgraph*}

\begin{fmfgraph*}(80,40)
\fmfkeep{uchannel}
\fmfleft{i1,i2}
\fmfright{o1,o2}
\fmf{plain}{i1,v1}
\fmf{phantom}{v1,o1} 
\fmf{plain}{i2,v2}
\fmf{phantom}{v2,o2} 
\fmf{plain,left=0.5,tension=0.4}{v1,v2}
\fmf{plain,right=0.5,tension=0.4}{v1,v2}
\fmf{plain,tension=0}{v1,o2}
\fmf{plain,tension=0}{v2,o1}
\fmfdot{v1,v2}
\end{fmfgraph*}

\begin{fmfgraph*}(80,40)
\fmfkeep{tadpolesix}
\fmfbottom{i1,o1}
\fmftop{i2,o2}
\fmf{plain}{i1,v1,o1}
\fmf{plain}{i2,v1,o2}
\fmf{plain,right=90,tension=0.8}{v1,v1}
\fmfdot{v1}
\end{fmfgraph*}


\begin{fmfgraph*}(100,36)
\fmfkeep{bubble}
\fmfleft{in}
\fmfright{out}
\fmfdot{v1}
\fmfdot{v2}
\fmf{plain}{in,v1}
\fmf{plain}{v2,out}
\fmf{plain,left,tension=0.6}{v1,v2}
\fmf{plain,right,tension=0.6}{v1,v2}
\end{fmfgraph*}


\begin{fmfgraph*}(100,36)
\fmfkeep{tadpole}
\fmfset{dash_len}{6pt} 
\fmfleft{in,p1}
\fmfright{out,p2}
\fmfdot{c}
\fmf{plain}{in,c}
\fmf{plain}{c,out}
\fmf{plain,right, tension=0.8}{c,c}
\fmf{phantom, tension=0.2}{p1,p2}
\end{fmfgraph*}

\end{fmffile}

}

\hfill{Imperial-TP-LW-2014-02}

\vspace{40pt}

\begin{center}
{\huge{\bf The one-loop worldsheet S-matrix {for the}
\vskip 6pt

 $\mathbf{AdS_n\times S^n\times T^{10-2n}}$  superstring}}

\vspace{20pt}

{\bf \large  Radu Roiban$^a$, Per Sundin$^b$, Arkady Tseytlin$^c{}^,$\footnote{Also at Lebedev Institute, Moscow}} and {\bf Linus Wulff$\, ^c$}

\vspace{15pt}

{$^a$ \it\small Department of Physics, The Pennsylvania State University, University Park,\\ Pennsylvania 16802, U.S.A.}\\
{$^b$ \it\small Universit\'a di Milano-Bicocca and INFN Sezione di Milano-Bicocca,\\ Dipartimento de Fisica,
Piazza della Scienza 3, I-20126 Milano, Italy}\\
{$^c$ \it\small The Blackett Laboratory, Imperial College, London SW7 2AZ, U.K.}\\

\vspace{50pt}

{\bf Abstract}

\end{center}
\noindent
We compute   the massive-sector 
 worldsheet S-matrix  
  for superstring theories  in $AdS_n\times S^n\times T^{10-2n}$   (with  $n=2,3,5$)
  in the near BMN  expansion   up to one-loop order in inverse string tension. 
We show that, after taking into account the wave function renormalization, the one-loop S-matrix
 is UV finite.
In an appropriate regularization scheme  the S-matrix 
 is   consistent with the  underlying symmetries   of the superstring  theory, 
 i.e.   for the $n=3,5$ cases it coincides with the one implied   by 
  the  light-cone gauge 
 symmetries   with  the dressing phases determined from the crossing equations. 
 For the $n=2,3$  cases  we  observe 
  that  the massless modes decouple
  from the one-loop calculation of  massive mode  scattering, i.e. 
    the $2n$-dimensional 
    supercoset sigma model and the full 10-dimensional   superstring happen to have the same  massive one-loop  S-matrix.

\vskip 90pt
\pagebreak
\setcounter{page}{1}


\section{Introduction}

The Green-Schwarz (GS) superstring   sigma model corresponding to a consistent 10d supergravity background 
should be one-loop UV finite when considered in conformal gauge  and restricted to on-shell 
values of background worldsheet  fields. This applies, in particular, to the one-loop partition function of the $AdS_5\times S^5$   
superstring  evaluated on a classical string solution.  Divergences  may appear  if one first  solves for the 2d metric (i.e. starts 
with Nambu-Goto-type action) or  considers off-shell correlators. 
Then  the computation of the worldsheet  S-matrix  near some vacuum such as the one provided by the BMN point-like string  (usually done in  a  ``light-cone"  or mixed coordinate-momentum gauge) 
 may not a priori   produce a UV finite result.
 
  Indeed, past  attempts of one-loop 
 BMN S-matrix   computations led to UV divergent results. 
 This is puzzling  as one  would like to provide a perturbative one-loop  check of  formal constructions   of   BMN  vacuum  S-matrices in $AdS_n\times S^n\times T^{10-2n}$
 theories\footnote{We focus on these particular examples as they are the simplest to analyze perturbatively, having for example no cubic interaction vertices. The integrability of the string in these backgrounds was pointed out in \cite{Bena:2003wd,Babichenko:2009dk,Sorokin:2011rr}.}  which are based on  symmetry  considerations   and  general properties 
  (integrability, unitarity, crossing)   and assume that the S-matrix   should be   UV  finite. 
  The aim of the  present paper is to resolve this problem   by showing that
  the  one-loop worldsheet S-matrix   computed directly from the superstring action and 
  properly defined to account   for non-trivial wave-function renormalization is indeed 
  UV finite.\footnote{A finite theory may still require infinite wave-function renormalization, a well known example being $\mathcal N=4$ super Yang-Mills theory in dimensional regularization, see, e.g.,  \cite{Velizhanin:2010vw} and references there.}

Previous work on  one-loop   BMN   S-matrices in $AdS_n\times S^n\times T^{10-2n}$ include:
\begin{itemize}
	\item The near flat space (NFS) limit  computations ($n=2,3,5$) \cite{Maldacena:2006rv,Klose:2007wq,Klose:2007rz,Rughoonauth:2012qd,Sundin:2012gc,Sundin:2013ypa,Murugan:2012mf,Abbott:2013kka}
	\item Constructions based on generalized unitarity ($n=3,5$) \cite{Bianchi:2013nra,Engelund:2013fja,Bianchi:2014rfa}
	\item Computations of some finite  BMN amplitudes  ($n=2,3$) \cite{Sundin:2014sfa,Abbott:2013kka}.
\end{itemize}

Below  we will present  the direct computation of the full near-BMN  2-particle S-matrix  not  relying on  truncations or assumptions.
 We will find that   the divergences which appear  at intermediate stages 
 may be  interpreted as wave-function renormalization of the bosons and 
 they cancel  in the S-matrix defined according to standard rules.
 Furthermore,   there  exists a symmetry-preserving regularization scheme
    in which the resulting finite S-matrix matches
   the (massive sector)  S-matrix found previously  
   from symmetries and crossing  considerations for $n=3,5$ theories.
    In the $n=2$ case the perturbative 
    S-matrix agrees with earlier calculations performed in \cite{Abbott:2013kka} and the recent   suggestion      to fix the S-matrix  using  symmetries and the Yang-Baxter equation \cite{Hoare:2014kma}.

\

Let us summarize our main results. We are interested in the S-matrix for scattering of massive excitations at one loop in the near-BMN expansion. A naive direct  calculation shows that  some of the  one-loop scattering amplitudes appear to diverge
\begin{eqnarray}
\mathcal A^{(naive)}(zz\rightarrow zz)=\mathrm{infinite}\,,\qquad
\mathcal A^{(naive)}(yz\rightarrow yz)=\mathrm{finite}\,,\qquad
\mathcal A^{(naive)}(yy\rightarrow yy)=\mathrm{infinite}\,.
\nonumber
\end{eqnarray}
Here  $z$ and $y$ denote  the $AdS_n$ and $S^n$  bosonic excitations 
respectively.  However, to properly define the amplitudes and S-matrix   one needs to take into account 
the field (or ``wave-function")  renormalization\footnote{If it were possible to argue that the $y^2z^2$ vertices should not  be renormalized, then the finiteness of $\mathcal A^{(naive)}(yz\rightarrow yz)$ would imply  that the renormalization factors of the 
$z$ and $y$ fields should obey $Z_z Z_y = finite$, i.e. that the corresponding 
one-loop divergences  should have opposite sign.}. The latter is computed from the (unrenormalized) one-loop  {\it off-shell}  two-point functions\footnote{The masses will be set to $1$ in our conventions but we keep them here for clarity.}
\begin{equation}
\langle zz\rangle=\frac{iZ_z}{p^2-m^2}+\mathcal O(g^{-2})\,,\qquad\langle yy\rangle=\frac{iZ_y}{p^2-m^2}+\mathcal O(g^{-2})\,.
\end{equation}
Explicit calculations show that the  wave-function renormalization factors are   given by
\begin{equation}
Z_z=1{+}\frac{1}{4\pi\hat g}\big(-\frac{2}{\epsilon}+\ldots\big)\,,\qquad Z_y=1{-}\frac{1}{4\pi\hat g}\big(-\frac{2}{\epsilon}+\ldots\big)\ , \qquad 
\hat g=\left\{
\begin{array}{cc}
g & \mbox{for}\ \  n=3,5 \\
2g & \mbox{for}\  \  n=2 
\end{array}
\right.
\label{eq:Zs}
\end{equation}
where $g$ is the string tension (the effective worldsheet coupling is $g^{-1}$).\footnote{We will carry out the calculations in light-cone gauge, which is the special 
case $a=1/2$ of the  interpolating $a$-gauge of  \cite{Arutyunov:2006gs}. In general, the renormalization factors $Z_y$ and $Z_z$ may depend on the gauge-fixing parameter $a$  and, because of absence of a $Z_2$ symmetry between 
$y$ and $z$ fields, are not expected to be related by simply changing the sign of the one-loop term.} The UV divergence comes from the tadpole integral
\begin{equation}
\label{eq:gamma-reg}
\int\frac{d^2k}{(2\pi)^2}\frac{1}{k^2-m^2}\to \frac{i}{4\pi}\big(-\frac{2}{\epsilon}+\gamma_E+\log\frac{m^2}{4\pi}\big)\,,
\end{equation}
where we evaluated the integral in dimensional regularization in $2-\epsilon$ dimensions. There is no independent mass renormalization, which is consistent with the BMN vacuum being 1/2 BPS.

Note that the bosonic  field renormalization  is of   opposite signs for the $AdS_n$ and the $S^n$ excitations. While this
may appear at odds with the non-manifest BMN vacuum symmetry, e.g. $[PSU(2|2)]^2$ for $n=5$, all that we can ask is that the S-matrix have this symmetry, which it does.\footnote{It is worth mentioning that, in theories in which symmetries are
not manifest or realized only on shell, fields
belonging to the same representation/multiplet
may still be renormalized differently
 without spoiling the symmetry.
An example is provided by ${\cal N}=4$
 super Yang-Mills theory where, in a component formulation, vector and scalar
fields have different renormalization factors \cite{Velizhanin:2010vw}.}
%
It is also interesting to observe that the results are universal in $n$ assuming 
that  in the $AdS_2$ case the string tension $g$ is effectively replaced by $2g$. 
A similar effect has been noticed  earlier at one \cite{Abbott:2013kka} and two \cite{Murugan:2012mf} loops.\footnote{It was slightly hidden there due to the fact that the string tension was called $g/2$ instead of $g$.}
The two-point function of the fermions turns out not to get renormalized at the one-loop order.


Taking this wave-function renormalization into account,  the scattering amplitudes 
are given by\footnote{For a  standard definition of renormalized S-matrix elements  see, e.g., \cite{ZinnJustin:2002ru}.}
\begin{align}
&\mathcal A(zz\rightarrow zz)=(\sqrt{Z_z})^4\mathcal A^{(naive)}(zz\rightarrow zz)\,,\qquad\nn\\
&\mathcal A(yz\rightarrow yz)=(\sqrt{Z_y})^2(\sqrt{Z_z})^2\mathcal A^{(naive)}(yz\rightarrow yz)\,,\qquad \label{1}\\
&\mathcal A(yy\rightarrow yy)=(\sqrt{Z_y})^4\mathcal A^{(naive)}(yy\rightarrow yy)\,,
\nonumber
\end{align}
and these are found  to be finite,  implying that no other (coupling or vertex) renormalizations are indeed required. 

Equivalently, given a field theory with quartic (and higher-point) interaction vertices, one 
may start with a Lagrangian with $Z$-factors introduced for all terms. 
Requiring the two-point functions to be finite  determines  the wave-function $Z$-factors as in (\ref{eq:Zs}). 
Next, requiring that the on-shell four-point function is finite determines the  $Z$-factors in front of the quartic coupling. 
In our case their divergent part is given by $Z_{\phi_1\phi_2\phi_3\phi_4}=\sqrt{Z_{\phi_1}Z_{\phi_2}Z_{\phi_3}Z_{\phi_4}}$.
This structure implies that there is in fact no genuine renormalization of the quartic couplings as this is controlled by 
the ratio $Z_{\phi_1\phi_2\phi_3\phi_4}/\sqrt{Z_{\phi_1}Z_{\phi_2}Z_{\phi_3}Z_{\phi_4}}$ which is finite.\footnote{This is 
consistent with the corresponding beta-function being zero since it is  determined in terms of the same ratio  
$Z_{\phi_1\phi_2\phi_3\phi_4}/\sqrt{Z_{\phi_1}Z_{\phi_2}Z_{\phi_3}Z_{\phi_4}}$.} 


While  this wave-function renormalization renders the S-matrix finite
 one still has to be 
  careful with how one regularizes the divergent integrals that appear in intermediate steps:
  the regularization should  be consistent with underlying symmetries.\footnote{Equivalently,  preservation of symmetries (including hidden ones related to integrability) 
  may require a particular choice of finite counterterms, see, e.g., \cite{deVega:1981ka,deVega:1982sh,Hoare:2010fb}
  for the complex Sine-Gordon theory example.
    }
A  naive approach  based on  computing   all integrals in dimensional regularization 
gives an S-matrix which differs from the one determined by the  symmetries, i.e. 
 this   regularization breaks (or at least gives a different realization of) the symmetries preserved by the BMN vacuum.
As we shall explain  below,  there is   
an improved regularization prescription   based on 
 first reducing the one-loop  integrals 
 to a  few divergent (tadpole) integrals by using algebraic identities  in $d=2$ 
 and then  computing   the latter   integrals  in dimensional regularization.
 This  regularization 
 scheme leads to the same  one-loop  S-matrix 
 as determined by the 
 symmetries and crossing equations for $AdS_5\times S^5$    ($n=5$)  (see, e.g., 
 \cite{Arutyunov:2009ga})
 and $AdS_3\times S^3 \times T^4$   ($n=3$)   \cite{Borsato:2014hja}
  theories. 
 In $AdS_2 \times S^2 \times T^6$   ($n=2$)  the result 
   is compatible with previous calculations    performed in \cite{Abbott:2013kka} and the recent derivation of the S-matrix from symmetries and the Yang-Baxter equation in \cite{Hoare:2014kma}. 
   
  This  regularization  prescription 
   is  therefore compatible with the symmetries of the BMN vacuum and with integrability,   at least up to one loop order.  It also has the
   interesting feature   that the massless modes present in the $n=2$ and $n=3$ cases decouple completely from the computation of the massive S-matrix, i.e. completely cancel out from internal lines of one-loop graphs. In that sense 
    the supercoset model  appears to be equivalent to the full superstring 
     as far as the massive one-loop  S-matrix is concerned.  This feature should no longer be true 
     at two-loop order      (see for example \cite{Murugan:2012mf}).

The outline of this paper is as follows.  In section 2  we shall describe 
the  general  structure   of the 10d superstring action  to quartic order in fermions.  In section 3 we shall specify to  the  case of $AdS_n \times S^n \times T^{10-2n}$   theories  and fix the light-cone gauge adapted to the BMN vacuum.  Section 4 describes our regularization procedure. The results for the one-loop massive sector S-matrix are  presented in section 5 with details in appendix  B. Appendix A contains some relations between one-loop integrals. 
In appendix C we comment on the  computation of the near BMN S-matrix and dispersion relation in conformal gauge.

\section{Superstring action}\label{GSactions}

\def \SS  {{\cal S}}
The Green-Schwarz superstring action can be expanded  in powers of  fermions
(here $g$ denotes the string tension)
\begin{equation}
S=g\int d^2\xi\,(\mathcal L^{(0)}+\mathcal L^{(2)}+\ldots)\,.
\end{equation}
In the $AdS_5\times S^5$ case  it is known to all orders in fermions \cite{Metsaev:1998it} due to the background being maximally supersymmetric. However,
 in a general 10d supergravity background it is only known to quartic order \cite{Wulff:2013kga}. 
 The purely bosonic terms in the Lagrangian are 
\begin{equation}\label{ac}
\mathcal L^{(0)}=\frac12\gamma^{ij}e_i{}^ae_j{}^b\eta_{ab}+\frac12\varepsilon^{ij}B^{(0)}_{ij}\,,\qquad\gamma^{ij}=\sqrt{-h}h^{ij}\,,
\end{equation}
where we denote the  bosonic vielbein {pulled back to the worldsheet } by $e_i{}^a $ $(a=0,\ldots,9; \ i=0,1)$ and $B_{ij}^{(0)}=e_i{}^ae_j{}^bB^{(0)}_{ab}$ is the lowest component in the  Grassmann parameter 
 $\Theta$-expansion of the NSNS two-form superfield $B$.
  For the terms involving fermions we will follow   \cite{Wulff:2013kga}   and 
  write the expressions appropriate to type IIA supergravity, i.e. $\Theta^{\underline\alpha}$ will be a 32-component Majorana spinor. At the end we will  
  describe how to get the type IIB expressions by performing some simple substitutions.

The terms quadratic in fermions take the form
\begin{equation}
\mathcal L^{(2)}=\frac{i}{2}e_i{}^a\,\Theta\Gamma_aK^{ij}\mathcal D_j\Theta\,,\qquad K^{ij}=\gamma^{ij}-\varepsilon^{ij}\Gamma_{11}\,,
\end{equation}
where
\begin{equation}
\label{eq:DbA}
\mathcal D\Theta=
\big(d-\frac{1}{4}\omega^{ab}\Gamma_{ab}+\frac{1}{8}e^aM_a\big)\Theta\,,\qquad M_a=H_{abc}\,\Gamma^{bc}\Gamma_{11}+\SS\Gamma_a\,.
\end{equation}
Here  $\omega^{ab}$ is the spin connection, $H=dB$ is the NSNS three-form field strength
and the RR fields enter the action through the bispinor\footnote{Here $\phi$ is the dilaton and we use the convention $F^{(n)} = \frac{1}{n!}dx^{m_n}\wedge\dots\wedge dx^{m_1}F_{m_1\dots m_n}$ for the form fields.}
\begin{equation}
\label{eq:SbA}
\SS=e^\phi\big(\frac12F^{(2)}_{ab}\Gamma^{ab}\Gamma_{11}+\frac{1}{4!}F^{(4)}_{abcd}\Gamma^{abcd}\big)\,.
\end{equation}

The quartic fermionic terms  are somewhat more complicated  \cite{Wulff:2013kga}
\begin{align}
\mathcal L^{(4)}=&
-\frac{1}{8}\Theta\Gamma^a\mathcal D_i\Theta\,\Theta\Gamma_aK^{ij}\mathcal D_j\Theta
+\frac{i}{24}e_i{}^a\,\Theta\Gamma_aK^{ij}\mathcal M\mathcal D_j\Theta
+\frac{i}{192}e_i{}^ae_j{}^b\,\Theta\Gamma_aK^{ij}(M+\tilde M)\SS\Gamma_b\Theta
\nonumber\\
&{}
+\frac{1}{192}e_i{}^ce_j{}^d\,\Theta\Gamma_c{}^{ab}K^{ij}\Theta\,(3\Theta\Gamma_dU_{ab}\Theta-2\Theta\Gamma_aU_{bd}\Theta)
\nonumber\\
&{}
-\frac{1}{192}e_i{}^ce_j{}^d\,\Theta\Gamma_c{}^{ab}\Gamma_{11}K^{ij}\Theta\,(3\Theta\Gamma_d\Gamma_{11}U_{ab}\Theta+2\Theta\Gamma_a\Gamma_{11}U_{bd}\Theta)\,.
\label{eq:L4}
\end{align}
Here $\tilde M=\Gamma_{11}M\Gamma_{11}$  and we defined 
\begin{align}
\mathcal M^\alpha{}_\beta=&{}
M^\alpha{}_\beta
+\tilde M^\alpha{}_\beta
+\frac{i}{8}(M^a\Theta)^\alpha\,(\Theta\Gamma_a)_\beta
-\frac{i}{32}(\Gamma^{ab}\Theta)^\alpha\,(\Theta\Gamma_aM_b)_\beta
-\frac{i}{32}(\Gamma^{ab}\Theta)^\alpha\,(C\Gamma_aM_b\Theta)_\beta, 
\nonumber\\
M^\alpha{}_\beta=&{}
\frac12\Theta T\Theta\,\delta^\alpha_\beta
-\frac12\Theta\Gamma_{11}T\Theta\,(\Gamma_{11})^\alpha{}_\beta
+\Theta^\alpha\, (CT\Theta)_\beta
+(\Gamma^aT\Theta)^\alpha\,(\Theta\Gamma_a)_\beta, 
\label{eq:M}\\
T=&{}\frac{i}{2}\nabla_a\phi\,\Gamma^a+\frac{i}{24}H_{abc}\,\Gamma^{abc}\Gamma_{11}+\frac{i}{16}\Gamma_a\SS\Gamma^a\,,
\label{eq:T}
\\
U_{ab}=&{}\frac{1}{4}\nabla_{[a}M_{b]}+\frac{1}{32}M_{[a}M_{b]}-\frac{1}{4}R_{ab}{}^{cd}\,\Gamma_{cd}\,.
\label{eq:Uab}
\end{align}
The dilatino equation is $T\xi=0$ while the integrability condition for the gravitino equation is $U_{ab}\xi=0$, where $\xi$ is a Killing spinor \cite{Wulff:2013kga,Wulff:2014kja}.

To find the corresponding type IIB string  expressions
the $32$-component Majorana spinor $\Theta^{\underline\alpha}$ should be replaced by a doublet of $16$-component Majorana-Weyl spinors
 {$\Theta^{\alpha1},\Theta^{\alpha2}$.} Similarly,  the $32\times32$ Dirac matrices are replaced by the $16\times16$ ones as follows
\begin{equation}
\Gamma_a\rightarrow\gamma_a\,,\qquad \qquad\Gamma_{11}\rightarrow\sigma^3\ , 
\end{equation}
with one exception: $\Gamma_{11}T\rightarrow-\sigma^3T$.
Finally, instead of the bispinor 
$\SS$ defined in (\ref{eq:SbA}) one should use the expression appropriate to the type IIB theory\footnote{Here $\sigma^n$ are Pauli matrices.
For more details and definitions of the gamma-matrices see \cite{Wulff:2013kga}. }
\begin{equation}
\label{eq:SbB}
\SS=-e^\phi\big({i\sigma^2}\gamma^a F^{(1)}_a+\frac{1}{3!}\sigma^1\gamma^{abc}F^{(3)}_{abc}+\frac{1}{2\cdot5!}{i\sigma^2}\gamma^{abcde} F^{(5)}_{abcde}\big) \,.
\end{equation}
With these replacements all the previous expressions apply also for the superstring in a type IIB supergravity background.

The superstring action simplifies in the cases we are considering in this paper 
as  all RR background fields are constant (and there is no NSNS flux, $H_{abc}=0$)\footnote{
 We follow the conventions of \cite{Wulff:2014kja}.
 Note that for the $AdS_2$ and $AdS_3$ case we give the 
 fluxes of  the type IIA solution.
 The corresponding type IIB solution is 
 obtained by T-duality 
 in a torus direction. 
 For the $AdS_5$ case the full superstring action is known 
 in the form of a supercoset model \cite{Metsaev:1998it}. This supercoset model coincides with the GS
 action described above up to quartic order in $\Theta$ provided the coset representative is chosen as 
 $g=e^{x^mP_m}e^{\Theta^\alpha Q_\alpha}$ \cite{Kallosh:1998zx}. Since we will need the $\Theta^6$-terms for the 
 one-loop S-matrix in the fermionic sector we will use the supercoset model for our calculations in this case.}
\begin{align}
AdS_5\times S^5:&\quad F^{(5)}=4e^{-\phi}(\Omega_{AdS_5}+\Omega_{S^5})\,,\nonumber\\
AdS_3\times S^3\times T^4:&\quad F^{(4)}=2e^{-\phi}dx^9\wedge (\Omega_{AdS_3}+\Omega_{S^3})\,,\label{eq:fluxes}\\ \nn
AdS_2\times S^2\times T^6:&\quad 
{
F^{(2)}=e^{-\phi}\Omega_{AdS_2}\,,\quad
F^{(4)}=-e^{-\phi}\Omega_{S^2}\wedge\big(dx^5\wedge dx^4+dx^7\wedge dx^6+dx^9\wedge dx^8\big)
}
\end{align}
Here the  $AdS_n$ and $S^n$ radii are set to be 1. We also  find from (\ref{eq:SbA}), (\ref{eq:SbB}) and (\ref{eq:T})
\begin{eqnarray}
\begin{array}{rll}
AdS_5\times S^5:&\quad \SS=-4{i\sigma^2}\gamma^{01234}\,,                &  T=0\,, \\[4pt]
AdS_3\times S^3\times T^4:&\quad \SS=-4\mathcal P_{16}\Gamma^{0129}\,, & T=-\displaystyle{\frac{i}{2}}\Gamma^{0129}(1-\mathcal P_{{16}})\,,\\[8pt]
AdS_2\times S^2\times T^6:&\quad 
{
\SS=-4\mathcal P_8\Gamma^{01}\Gamma_{11}}\,, & {T=\displaystyle{\frac{i}{2}}\Gamma^{01}\Gamma_{11}(1-\mathcal P_8)}
\,,
 \end{array}
\end{eqnarray}
where we have defined the following three projection operators{, with the dimension of the subspace they project on, i.e. the number of supersymmetries preserved by the background, indicated}
\begin{equation}
{
\mathcal P_{16}=\frac12(1+\Gamma^{012345})\,,\qquad
\mathcal P_8=\frac14(1-\Gamma^{4567}-\Gamma^{4589}-\Gamma^{6789})\,.
}
\end{equation}
We  will take the metric of $AdS_n $ in the  form
\begin{equation}
ds^2_{AdS_n}=-\Big(\frac{1+\frac12|z_I|^2}{1-\frac12|z_I|^2}\Big)^2dt^2+\frac{2|dz_I|^2}{(1-\frac12|z_I|^2)^2}\qquad I=1,\ldots,( n-1)/2\,,
\label{eq:AdSmetric}
\end{equation}
where the spatial  coordinates are grouped together into two complex coordinates in $AdS_5$, one in $AdS_3$ and one real coordinate $x_1=\sqrt2z$ in $AdS_2$. Similarly,  the $S^n$ metric is
\begin{equation}
ds^2_{S^n}=\Big(\frac{1-\frac12|y_I|^2}{1+\frac12|y_I|^2}\Big)^2d\varphi^2+\frac{2|dy_I|^2}{(1+\frac12|y_I|^2)^2}\qquad I=1,\ldots,(n-1)/2\,.
\label{eq:Smetric}
\end{equation}
Again,  we use $x_2=\sqrt2y$ for the real coordinate in $S^2$.


\section{Near BMN expansion of the $AdS_n\times S^n\times T^{10-2n}$
string action}\label{sec:BMN-expansion}

Given the form of the background fields we can now expand the string 
action around the BMN  vacuum  
$t=\varphi=\tau$  \cite{Berenstein:2002jq}.
We shall  fix the light-cone gauge and the corresponding kappa symmetry gauge as
\begin{equation}
x^+\equiv \frac12(t+\varphi) =\tau\,,\qquad\Gamma^+\Theta=0\,, 
\end{equation}
where the  complete  gauge fixing also includes the conditions 
\begin{equation}
p^+\equiv -\frac12\frac{\partial\mathcal L}{\partial \dot x^-}=1\,,\qquad\frac{\partial\mathcal L}{\partial x^-{}'}=0\,.
\end{equation}
In this gauge the worldsheet metric in (2.2) 
   takes the form $\gamma_{ij}=\eta_{ij}+\hat\gamma_{ij}$, where $\hat\gamma_{ij}$ denotes higher order corrections to be determined from the  above conditions.\footnote{The Virasoro constraints can be   used to solve for $x^-$ whose explicit form we will not need here.} 

Next, let us consider the near BMN  expansion of the action, 
i.e. in powers of the transverse coordinates and fermions.
 Scaling all transverse fields  
 with a factor $g^{-1/2}$ yields 
\bea \nn
\mathcal{L}= \mathcal{L}_2+\frac{1}{g}\mathcal{L}_4+\frac{1}{g^2}\mathcal{L}_6+\dots\ , 
\eea
where the subscript denotes the number of transverse coordinates in each term. Note that only terms with an even number of fields appear in the expansion. This fact  simplifies  the perturbative expansion in the $AdS_n \times S^n $  case   compared to more complicated backgrounds.
 The quadratic Lagrangian $\mathcal{L}_2$ takes the form\footnote{Here $\partial_\pm=\partial_0\pm\partial_1$ and massless modes have a primed index.}
\begin{align}
\label{eq:quadratic-L}
\mathcal{L}_2=&
|\partial_iz_I|^2
-|z_I|^2
+|\partial_iy_I|^2
-|y_I|^2
+|\partial_iu_{I'}|^2 
+i\bar\chi_L^r \partial_-\chi_L^r
+i\bar\chi_R^r \partial_+\chi_R^r
-\bar\chi_L^r\chi_R^r
-\bar\chi_R^r\chi_L^r
\nonumber\\
&{}
+i\bar\chi_L^{r'}\partial_- \chi_L^{r'}
+i\bar\chi_R^{r'}\partial_+\chi_R^{r'}\,.
\end{align}
The field content of the $n=5,3,2$  theories  is summarized in table \ref{tab:1} and the $U(1)$ charges are summarized in tables \ref{tab:charges-ads5}--\ref{tab:charges-ads2}. For the interaction terms we will only give the bosonic terms quartic in fields due to the length of the expressions: 
\begin{align}
\mathcal L_4^B=&
\frac12(|y_I|^2-|z_I|^2)\left(
|\partial_0z_I|^2
+|\partial_1z_I|^2
+|\partial_0y_I|^2
+|\partial_1y_I|^2
+|\partial_0u_{I'}|^2
+|\partial_1u_{I'}|^2
\right)
\nonumber\\
&{}
+|z_I|^2|\partial_iz_I|^2
-|y_I|^2|\partial_iy_I|^2
\,.
\label{eq:L4B}
\end{align}

\begin{table}[ht]
\begin{center}
\begin{tabular}{c|ccc|ccc}
 & & $m=1$ &  &  & $m=0$ & \\
\hline 
 & AdS & Sphere & Fermions & Torus &  & Fermions\\
\hline
\vphantom{${}^{A^{A^a}}$}
$AdS_5\times S^5$ & $z_1,z_2$ & $y_1,y_2$ & $\chi^{1,2,3,4}$ & -- &  & --\\
\vphantom{${}^{A^{A^A}}$}
$AdS_3\times S^3\times T^4$ & $z_1$ & $y_1$ & $\chi^{1,2}$ & $u_1,u_2$ & &$\chi^{3,4}$ \\
\vphantom{${}^{A^{A}}$}
$AdS_2\times S^2\times T^6$ & $x_1$ & $x_2$ & $\chi^1$ & $u_1,u_2,u_3$ & &$\chi^{2,3,4}$ \\
\hline
\end{tabular}
\end{center}
\caption{Summary of the field content. All fields are complex except $(x_1,\,x_2)=\sqrt2(z,\,y)$ in the $AdS_2\times  S^2\times T^6$ case. The massive fields ($m=1$) come from the supercoset model while the massless ones ($m=0$) are only present in the full  10d superstring theory.}
\label{tab:1}
\end{table}
\begin{table}[ht]
\begin{center}
\begin{tabular}{ccccccccc}
& $y_1$ & $y_2$ & $z_1$ & $z_2$ & $\chi^1$ & $\chi^2$ & $\chi^3$ & $\chi^4$ \\
\hline
\vphantom{${}^{A^{A^a}}$}
$U(1)_1$ & $-1$ & $0$ & $0$ & $0$ & $-1/2$ & $1/2$ & $1/2$ & $1/2$  \\
$U(1)_2$ & $0$ & $-1$ & $0$ & $0$ & $1/2$ & $-1/2$ & $1/2$ & $1/2$ \\
$U(1)_3$ & $0$ & $0$ & $-1$ & $0$ & $1/2$ & $1/2$ & $-1/2$ & $1/2$  \\
$U(1)_4$ & $0$ & $0$ & $0$ & $-1$ & $1/2$ & $1/2$ & $1/2$ & $-1/2$
\end{tabular}
\end{center}
\caption{Summary of $U(1)$ charges for $AdS_5\times S^5$.}
\label{tab:charges-ads5}
\end{table}
\begin{table}[ht]
\begin{center}
\begin{tabular}{ccccccccc}
& $y_1$ & $z_1$ & $u_1$ & $u_2$ & $\chi^1$ & $\chi^2$ & $\chi^3$ & $\chi^4$ \\
\hline
\vphantom{${}^{A^{A^a}}$}
$U(1)_1$ & $-1$ & $0$ & $0$ & $0$ & $-1/2$ & $1/2$ & $1/2$ & $1/2$  \\
$U(1)_2$ & $0$ & $-1$ & $0$ & $0$ & $1/2$ & $-1/2$ & $1/2$ & $1/2$ \\
$U(1)_3$ & $0$ & $0$ & $-1$ & $0$ & $1/2$ & $1/2$ & $-1/2$ & $1/2$
\end{tabular}
\end{center}
\caption{Summary of $U(1)$ charges for $AdS_3\times S^3\times T^4$. The $U(1)$'s associated to $T^4$    are  compatible with the fluxes in (\ref{eq:fluxes})  assuming  $u_1=\frac{1}{\sqrt2}(x^6+ix^7)$ and $u_2=\frac{1}{\sqrt2}(x^8+ix^9)$.}
\label{tab:charges-ads3}
\end{table}
\begin{table}[ht]
\begin{center}
\begin{tabular}{ccccccccc}
& $x_1,x_2$ & $u_1$ & $u_2$ & $u_3$ & $\chi^1$ & $\chi^2$ & $\chi^3$ & $\chi^4$ \\
\hline
$U(1)_1$ & $0$ & $-1$ & $0$ & $0$ & $-1/2$ & $-1/2$ & $1/2$ & $1/2$  \\
$U(1)_2$ & $0$ & $0$ & $-1$ & $0$ & $-1/2$ & $1/2$ & $-1/2$ & $1/2$ \\
$U(1)_3$ & $0$ & $0$ & $0$ & $-1$ & $-1/2$ & $1/2$ & $1/2$ & $-1/2$
\end{tabular}
\end{center}
\caption{Summary of $U(1)$ charges for $AdS_2\times S^2\times T^6$. The
 $U(1)$'s associated to $T^6$    are  compatible with the fluxes in (\ref{eq:fluxes})  assuming 
  $u_1=\frac{1}{\sqrt2}(x^4+ix^5)$, $u_2=\frac{1}{\sqrt2}(x^6+ix^7)$ and $u_3=\frac{1}{\sqrt2}(x^8+ix^9)$.}
\label{tab:charges-ads2}
\end{table}

\section{Regularization procedure}\label{sec:regularization}

For the $AdS_n\times S^n \times T^{10-2n} $  backgrounds under consideration the string 
Lagrangian  expanded near the BMN vacuum contains fourth and sixth order interaction vertices.
The one-loop contribution to the two-point function comes from tadpole diagrams with topology
\bea
\label{tadpole}
\parbox[top][0.95in][c]{1.5in}{\fmfreuse{tadpole}}
\eea
while the contribution to the four-point function comes from the three ($s,t$ and $u$-channel) bubble diagrams 
\bea
\label{bubble-diagrams}
\parbox[top][0.8in][c]{1.5in}{\fmfreuse{schannel}}
\quad + \quad
\parbox[top][0.8in][c]{1.5in}{\fmfreuse{tchannel}}
\quad + \quad 
\parbox[top][0.8in][c]{1.5in}{\fmfreuse{uchannel}}
\eea
and one tadpole diagram arising from the sixth order  interaction  term
\vspace{.5cm}
\bea
\label{tadpole-diagrams}
\parbox[top][0.8in][c]{1.5in}{\fmfreuse{tadpolesix}}
\eea
We will now describe how we evaluate these.

In the calculation of the one-loop Feynman diagrams involving only massive fields one encounters the following bubble integrals, corresponding to the diagrams in eq.~(\ref{bubble-diagrams}),
\begin{equation}
B^{r,s}(P)\equiv \int\frac{d^2k}{(2\pi)^2}\,\frac{k_+^rk_-^s}{(k^2-m^2)((k-P)^2-m^2)}\,,
\label{eq:bubbles}
\end{equation}
where $P$ is a combination of the external momenta, and  also the 
tadpole integrals  (corresponding to  eq.~(\ref{tadpole-diagrams}))
\bea
T^{r,s}(P)\equiv \int\frac{d^2k}{(2\pi)^2}\,\frac{k_+^rk_-^s}{(k-P)^2-m^2}\,.
\label{eq:tadpoles}
\eea
Many of these integrals are UV-divergent and need to be regularized. 
For the two-point function determining the wave function 
 renormalization  we only have a tadpole contribution in eq.~(\ref{tadpole})
  which we simply evaluate in dimensional regularization.

Given a sum of loop integrals one has several options to evaluate it. 
One may simply introduce Feynman parameters 
and evaluate the integrals one by one in,  e.g.,  dimensional regularization. In the presence of power-like divergences this is typically
dangerous as dimensional regularization amounts to an uncontrolled subtraction of such divergences which may include 
finite terms as well.
A safer alternative is to employ the reduction to master integrals. In this approach one uses algebraic identities as well 
as identities valid only after integration to express the original regularized integrals as linear combinations of a smaller 
set of integrals which are in some sense linearly independent (e.g. they do not have overlapping branch cuts).
For the same reason as before, use of integral identities for dimensionally-regulated power-divergent   
integrals may lead to an uncontrolled elimination of finite terms with rational momentum dependence.
Here we will use a variant of this approach which makes use of only algebraic identities and is {similar in spirit to} what 
is  sometimes called ``implicit regularization". It proceeds in  the following steps:\footnote{A similar procedure was used in \cite{Sundin:2014sfa} but tadpoles were written in terms of bubbles instead of the other way around. We have checked that our  present procedure does not change any of the results obtained there.}
\begin{itemize}
	\item[1.] Use algebraic identities on the integrands to reduce  the result 
	 to a minimal set of divergent integrals, in our case tadpole integrals.\footnote{There is typically no standard choice for this set of integrals. One simply has to find (if possible) a set which leads to a result compatible with the symmetries one wants to preserve.   Note   also that 
	  we do not  allow shifts of loop variables as this can be problematic in divergent integrals.}
	\item[2.] Evaluate these in a suitable regularization scheme  consistent with the algebraic identities used in the first step and the symmetries we want to preserve; 
	 in our case   this regularization is dimensional regularization.
\end{itemize}
Let us now apply this procedure to the integrals appearing in the  problem of  two-particle  scattering.   The first step will be to use the identity $k_+k_-=k^2-m^2+m^2$, which implies
\begin{equation}
B^{r,s}(P)=T^{r-1,s-1}(P)+m^2B^{r-1,s-1}(P)\,.
\label{eq:bubbles-id1}
\end{equation}
This allows us to reduce all relevant bubble integrals  to the following  set
\begin{equation}
B^{r,0}(P)\,,\qquad B^{0,s}(P)\,,\qquad r,s=0,1,2,3\,.
\end{equation}
These are still (potentially) divergent\footnote{In a Lorentz-invariant regularization scheme such as dimensional regularization they are finite. Here,  however, we are not interested in preserving Lorentz invariance but rather  a  non-relativistic symmetry  of the BMN vacuum.}
 for $r,s \ge 2$ so we want to reduce them further. This can be done by using the identity
\begin{equation}
\frac{1}{(k-P)^2-m^2}=\frac{1}{k^2-m^2}+\frac{2k\cdot P-P^2}{((k-P)^2-m^2)(k^2-m^2)}\,,
\end{equation}
which implies
\begin{equation}
P_-B^{r+1,s}(P)+P_+B^{r,s+1}(P)=T^{r,s}(P)-T^{r,s}(0)+P^2B^{r,s}(P)\,.
\label{eq:bubbles-id2}
\end{equation}
Combining this with the previous identity (\ref{eq:bubbles-id1}) we can reduce all bubble integrals to $B^{00}$ and $B^{01}$, which are finite without any regularization, and tadpole integrals. 

So far we have used only algebraic identities and made no shifts in loop variables. The next step is to note that since $B^{01}$ is finite we are 
 allowed to shift the integration variable. Making the shift $k\rightarrow-k+P$ gives
\begin{equation}
 B^{01}(P)=\frac12P_-B^{00}(P)\,.
\end{equation}
We now note the important fact that for this identity to be consistent with eq. (\ref{eq:bubbles-id2}) for $(r,s)=(0,0)$ we must have
\begin{equation}
T^{00}(P)-T^{00}(0)=0\,,
\label{eq:tadpole-shift}
\end{equation}
i.e. we should be allowed to shift the loop variable in the $T^{00}$ tadpole integral. 
It is then consistent to also allow shifts of loop variables in 
 other tadpole integrals\footnote{We could of course instead just compute them directly in dimensional regularization.} which reduces them further to an even smaller set. 
 
 In the end we are left  only with  $B^{00}(P)$, $T^{11}(0)$ and $T^{00}(0)$ (see appendix \ref{sec:int-reduction}). The two  tadpole integrals   $T^{11}(0)$ and $T^{00}(0)$ 
 can be computed  in dimensional regularization which respects (\ref{eq:tadpole-shift}) and the remaining  bubble integral  $B^{00}(P)$
  is manifestly finite.\footnote{One  reason for using dimensional regularization in this 
   last step is that  it removes the quadratic divergence in  $T^{11}(0)$. This quadratic divergence  appears not to be consistent with the symmetries of the BMN vacuum. In general, 
   there may be  additional quadratically divergent terms   
    coming   from the measure and local field redefinition factors,   and  use of dimensional regularization    allows us to ignore them too.}

Let us note  that the 
 fact that computing all integrals in dimensional regularization (without using any algebraic identities) gives a different answer can be seen by looking,  for example, 
  at $T^{01}(P)$ which is linearly divergent. 
  In dimensional regularization we can shift the loop variable to get 
\begin{equation}
T^{01}(P)=T^{01}(0)+P_-T^{00}(0)=P_-T^{00}(0)\,.
\end{equation}
On the other hand,  we could use the algebraic identities (\ref{eq:bubbles-id2}) and (\ref{eq:bubbles-id1}) to write
\begin{equation}
T^{01}(P)
=P_-T^{00}(P)+P_+B^{02}(P)+(m^2-\frac12P^2)P_-B^{00}(P)\,.
\end{equation}
The right hand sides in these two expressions are not equal in dimensional regularization --   they differ by rational terms coming from $B^{02}(P)$.\footnote{The integral $B^{02}(P)$ contains a divergence which happens to be a total derivative. In dimensional regularization this term gives no contribution but in a regularization which keeps surface terms it will contribute additional rational terms. This is the origin of the regularization ambiguity.} 
We find it  natural to require  that algebraic identities should always 
 hold and only allow shifts in loop momenta when it is consistent with this requirement.

In the $AdS_3\times S^3 \times T^4 $ and $AdS_2\times S^2 \times T^6 $ cases
 we also have massless modes in the near-BMN action 
  which means that we will have integrals of the form (\ref{eq:bubbles}) and (\ref{eq:tadpoles}) with $m=0$. Note that no bubble integrals involving different masses appear in  one-loop  diagrams  contributing to the two-particle S-matrix with all massive 
  external states\footnote{Bubble integrals with one massive and one massless internal propagator appear in the calculation 
  of the two-point function of massive fields in conformal gauge discussed in Appendix~\ref{app:conformal2pf}.}.
  The integrals which appear can be reduced in the same way as described above 
  (using essentially the same identities). 
   We shall  treat IR-divergent integrals by introducing a small regulator mass. In the end it turns out that the massless modes give no contribution and thus could be truncated away from the beginning, i.e. the supercoset sigma model gives the full answer for the massive 
   S-matrix even though it is not in general equivalent to the full superstring theory 
   (at least not in the $AdS_2\times S^2 \times T^6 $   case where the supercoset model  cannot  be obtained by kappa symmetry gauge-fixing of the full 10d superstring action though it is a consistent classical truncation \cite{Sorokin:2011rr}). 
   
   This decoupling of the massless modes only 
    holds in the regularization described  above, i.e. is not true in general. 
    For example, if one computes the S-matrix in the near-flat-space limit 
    one gets the correct result by just using dimensional regularization but in that regularization
     the massless modes give a non-vanishing contribution (see,  for example, 
      \cite{Abbott:2013kka}). 
     If one used the regularization described  above   one would find again
     that they decouple, with the final result still being the same.

\section{One-loop massive sector S-matrix}
\label{sec:scattering}

Having established the notation and the regularization scheme
 we now turn to the perturbative  computation of the worldsheet S-matrix,
\bea
\label{eq:S-matrix-action}
\mathbbm{S} = \mathbbm{1} + i \mathbbm{T},\qquad \mathbbm{T}=\frac{1}{g}\mathbbm{T}^{(0)} + \frac{1}{g^2}\mathbbm{T}^{(1)}+\mathcal{O}(g^{-3})\ , 
\eea
where the superscripts $(0)$ and $(1)$ denote the tree-level and one-loop contributions, respectively. 
The $\mathbbm T$-matrix maps a two-particle in-state to a corresponding two-particle out-state 
\bea
\label{eq:T-matrix-action}
\mathbbm{T}  | A(p) B(q) \rangle = T^{CD}_{AB}  | C(p) D(q) \rangle\ , 
\eea
where the capital letters denote any type of bosonic or fermionic excitation. We will 
ignore  the imaginary terms in $T_{AB}^{CD}$ since they are completely determined in terms of tree-level amplitudes via the optical theorem and  are not sensitive to regularization.
 Furthermore, for an integrable system in two dimensions the 
 energy-momentum conservation implies that the outgoing momenta are at most a permutation of the incoming momenta $p$ and $q$.

The specific in-  and out-states that we will consider consist of massive 
bosonic and fermionic excitations. For the $n=5$ or  $n=3$  theories
 where the worldsheet fields are complex, we will denote the two-particle asymptotic states as{
\begin{align}
\ket{z_\pm^I(p) z_\pm^J (q)}\,,\quad\ket{z_\pm^I(p) y_\pm^J (q) }\,,\quad \ket{z_\pm^I (p)\chi_\pm^r(q)}\,,\quad \ket{y_\pm^I(p) \chi_\pm^r(q)}\,,\quad\ket{\chi_\pm^r(p)\chi^s_\pm(q)}\,,
\end{align}
where $r,s=1,...,4$  or $r,s=1,2$, \ \  $I,J=1,2$  or $I,J=1$
 and the $\pm$ subscript refers to the $U(1)$ charge of a particle (see tables \ref{tab:charges-ads5} and \ref{tab:charges-ads3}). For the $n=2$ theory, on the other hand, we have real bosons and the relevant states will be denoted  as 
\bea
\ket{ x^k(p) x^l(q)}\,,\quad\ket{x^k(p) \chi_\pm^1(q)}\,,\quad\ket{\chi^1_\pm(p)\chi^1_\pm(q)}\ , \qquad \ \ \ k,l=1,2\,.
\eea}
Having set up the notation let us  now present the results 
of the  computations. We will start with 
the $n=5$ case where we will first  compute  the amplitudes directly,  without implementing the wave function renormalization (\ref{eq:Zs}),   and then show  how 
the UV divergences cancel in the properly defined  S-matrix elements \rf{1}. 

\subsection{$AdS_5 \times S^5$}
Let us  start   with  processes where we scatter $z$ and $y$ particles separately. Evaluating the amplitude, which is given by a sum of the topologies (\ref{bubble-diagrams}) and (\ref{tadpole-diagrams})  we get 
\bea
&&\mathbbm{T} \ket{z_\pm^I(p) z_\pm^I(q)}=\ell_1^z \ket{z_\pm^I(p) z_\pm^I(q)},\qquad \mathbbm{T} \ket{y_\pm^I(p) y_\pm^I(q)}=\ell_1^y \ket{y_\pm^I(p) y_\pm^I(q)} \ , 
\\
&& 
\ell_1^z = -\frac{1}{g} l_1 + \frac{1}{g^2}\Big( 2\Theta_{HL}+ \frac{1}{2\pi} \gamma(\epsilon)l_1\Big),\qquad
\ell_1^y = \frac{1}{g} l_1 + \frac{1}{g^2}\Big( 2\Theta_{HL}+ \frac{1}{2\pi} \gamma(\epsilon)l_1\Big),
\eea
where $l_1$ is the corresponding tree-level amplitude, $\Theta_{HL}$ is the one-loop 
contribution  corresponding to the well known Hernandez-Lopez phase \cite{Hernandez:2006tk, Beisert:2006ib} and
\bea
\label{eq:gamma}
\gamma(\epsilon)=-\frac{2}{\epsilon}+\gamma_E-\log 4\pi\ . 
\eea
The  terms with $\gamma(\epsilon)$ are arising from the integral (\ref{eq:gamma-reg}) evaluated in dimensional regularization. For the explicit representation of the HL phase  term 
in our conventions  see (\ref{eq:HL-phase}). As was mentioned  above, we are  ignoring imaginary terms in $\ell_1^z,\ell_1^y$.

Implementing  the wave-function renormalization (\ref{eq:Zs}) we see that the above amplitudes become finite.  At the same time, 
the  scattering amplitude mixing equal numbers of $z$ and $y$ particles 
also remains finite as  the contributions from the wave-function renormalization cancel each other out. Indeed, we find 
\bea
\mathbbm{T} \ket{z_\pm^I(p) y_\pm^J(q)}= \ell_2\ket{z_\pm^I(p) y_\pm^J(q)}{+\mbox{fermions}}\,,\qquad
\ell_2=-\frac{1}{g}l_2+\frac{1}{g^2}2\Theta_{HL} \ . 
\eea
For the  scattering amplitudes involving two fermions in the final state we get
\bea  &&
\mathbbm{T} \ket{z_\pm^I(p) z_\mp^I(q)}=\sum_{r=1}^4 \ell_{3,z}^r \ket{\chi_\pm^r(p)\chi_\mp^r(q)}+\ldots \ , \\ &&
\mathbbm{T} \ket{y_\pm^I(p) y_\mp^I(q)}=\sum_{r=1}^4 \ell_{3,y}^r \ket{\chi_\pm^r(p)\chi_\mp^r(q)}+\ldots\ , \\
&&
\ell^r_{3,z}=\Big(-\frac{1}{g}l_3+\frac{1}{g^2}\frac{1}{4\pi}\gamma(\epsilon)l_3\Big)\delta_{I+2,r}\,,\qquad
\ell^r_{3,y}=\Big(\frac{1}{g}l_3+\frac{1}{g^2}\frac{1}{4\pi}\gamma(\epsilon)l_3\Big)\delta_{Ir}\,. 
\eea
The fermions should not be renormalized  (as implied  by  the off-shell finiteness of the two-point functions of fermions), 
and  taking  into account the wave-function renormalization of the bosons 
the corresponding  S-matrix elements   become finite. 

In order to  provide a further consistency check of our 
  regularization   method, let us consider a few more amplitudes. For example, for the diagonal scattering				
\bea
\nn
\mathbbm{T} \ket{z_\pm^I(p) \chi_\pm^r(q)}_{I+2=r}\!\!\!&=&\!\!\!\ell^{z\chi}_4\ket{z_\pm^I(p) \chi_\pm^r(q)}+\dots,\qquad
\mathbbm{T} \ket{y_\pm^I(p) \chi_\pm^r(q)}_{I=r}=\ell^{y\chi}_4\ket{y_\pm^I(p) \chi_\pm^r(q)}+\dots,\\ \nn
\mathbbm{T} \ket{z_\pm^I(p) \chi_\pm^r(q)}_{I+2\neq r}\!\!\!&=&\!\!\!\ell^{z\chi}_5\ket{z_\pm^I(p) \chi_\pm^r(q)}+\dots,\qquad
\mathbbm{T} \ket{y_\pm^I(p) \chi_\pm^r(q)}_{I\neq r}=\ell^{y\chi}_5\ket{y_\pm^I(p) \chi_\pm^r(q)}+\dots,
\eea
we find
\bea 
&& \ell^{z\chi}_4=-\frac{1}{g} l_4+\frac{1}{g^2}\Big(2\Theta_{HL}+\frac{1}{4\pi} \gamma(\epsilon) l_4\Big),\qquad
\ell^{y\chi}_4=\frac{1}{g} l_4+\frac{1}{g^2}\Big(2\Theta_{HL}+\frac{1}{4\pi} \gamma(\epsilon) l_4\Big),\\ 
&&
\ell^{z\chi}_5=-\frac{1}{g} l_5+\frac{1}{g^2}\Big(2\Theta_{HL}+\frac{1}{4\pi} \gamma(\epsilon) l_5\Big),\qquad
\ell^{y\chi}_5=\frac{1}{g} l_5+\frac{1}{g^2}\Big(2\Theta_{HL}+\frac{1}{4\pi} \gamma(\epsilon) l_5\Big)\ , 
\eea
which again  correspond to finite   S-matrix elements  after renormalization of only bosonic legs. 

This also  implies that all  two-fermion scattering amplitudes   should be finite at one-loop level. To check  this explicitly the superstring  action to sixth order in fermions is needed. 
In the $AdS_5\times S^5$ the  action is  given by the supercoset  construction to all orders in fermions   (see 
e.g. \cite{Arutyunov:2009ga}). After some work we  indeed find  a finite  result
\bea
\mathbbm{T} \ket{\chi_\pm^r(p) \chi_\pm^s(q)}&=&\ell_6^{rs}\ket{\chi_\pm^r(p) \chi_\pm^s(q)}+\dots,\qquad
\ell_6^{rs}=-\frac{1}{g} l_6(1-\delta_{rs})+ \frac{2}{g^2} \Theta_{HL} \ .
\eea
Here $r,s=1,2$ or $3,4$ and the Kronecker delta indicates that only scattering with different fermionic flavors 
have a non-zero tree-level term.\footnote{For scattering processes with $r=1,2$ and $s=3,4$  the tree-level amplitudes vanish identically.} 

To summarize, taking the wave-function renormalization (\ref{eq:Zs}) into account
the one-loop contributions  to the diagonal S-matrix elements
 are finite and completely captured by the HL  phase term:
\bea
\label{eq:ads5-amplitudes} \nn
&& 
\ell_1^z = -\frac{1}{g} l_1 + \frac{2}{g^2}\Theta_{HL},\qquad
\ell_1^y = \frac{1}{g} l_1 + \frac{2}{g^2}\Theta_{HL}, \qquad \ell_2=-\frac{1}{g}l_2+\frac{2}{g^2}\Theta_{HL} \\ \nn
&&\ell^{z\chi}_4=-\frac{1}{g} l_4+\frac{2}{g^2}\Theta_{HL}, \qquad \ell^{y\chi}_4=\frac{1}{g} l_4+\frac{2}{g^2}\Theta_{HL}, \qquad
\ell^{z\chi}_5=-\frac{1}{g} l_5+\frac{2}{g^2}\Theta_{HL},
\\ 
&& 
\ell^{y\chi}_5=\frac{1}{g} l_5+\frac{2}{g^2}\Theta_{HL}, \qquad \ell_6^{rs}=-\frac{1}{g} l_6(1-\delta_{rs})+ \frac{2}{g^2}\Theta_{HL} \ . 
\eea
Additional imaginary parts of S-matrix elements, which as mentioned in the beginning of sec.~\ref{sec:scattering} we 
ignored in our calculation, may be restored through the optical theorem. 
Also, the renormalized off-diagonal  elements are 
\bea
&& \ell^r_{3,z}=-\frac{1}{g}l_3\delta_{I+2,r}\ ,\qquad
\ell^r_{3,y}=\frac{1}{g}l_3 \delta_{Ir}\ . \qquad 
\eea
The off-diagonal amplitudes are finite and the one-loop contribution is purely imaginary, i.e. fully determined via unitarity 
  by tree-level amplitudes.
All the resulting amplitudes are in complete agreement with the predictions (see, e.g., \cite{Arutyunov:2009ga})\footnote{Their $su(2|2)^2$-covariant fields are related to ours as follows:
\[
Z^{\alpha\dot\alpha}=
\frac{1}{\sqrt2}\left(\begin{array}{cc}
z^1_+ & z^2_-\\
-z^2_+ & z^1_-
\end{array}\right)\,,\quad
Y^{a\dot a}=
\frac{1}{\sqrt2}\left(\begin{array}{cc}
y^1_+ & y^2_-\\
-y^2_+ & y^1_-
\end{array}\right)\,,\quad
\]
\[
\eta^{\alpha\dot a}=
\frac{1-i}{2}\left(\begin{array}{cc}
\chi^1_R+\chi^1_L & -\chi^2_R-\chi^2_L\\
\bar\chi^2_R-\bar\chi^2_L & \bar\chi^1_R-\bar\chi^1_L
\end{array}\right)\,,\quad
\theta^{a\dot\alpha}=
\frac{1-i}{2}\left(\begin{array}{cc}
\chi^3_R+\chi^3_L & \bar\chi^4_R-\bar\chi^4_L \\
-\chi^4_R-\chi^4_L & \bar\chi^3_R-\bar\chi^3_L
\end{array}\right)\,,
\]
as can be seen by matching the $U(1)$-charges and comparing the quadratic terms in the action. Note that the requirement that our fermions have a standard kinetic term breaks the $su(2|2)^2$-covariance and causes the S-matrix elements involving fermions to take a slightly different form then in \cite{Arutyunov:2009ga}.}
 coming from symmetries and  integrability. 


\subsection{$AdS_3\times S^3\times T^4$}
For $AdS_3\times S^3\times T^4$ we will for simplicity restrict
consideration  to purely bosonic in- and out-states and we will implement the wave-function renormalization (\ref{eq:Zs}) from the  start.  Here we have, in total, 
 one  transverse (complex) boson in $AdS_3$ and  one in $S^3$. Looking at processes not mixing the two  we get\footnote{These amplitudes diverge before field renormalization.}
\bea \nn
&& \mathbbm{T} \ket{z_\pm^1(p) z_\pm^1(q)} = \ell^z_1  \ket{z_\pm^1(p) z_\pm^1(q)},\qquad \mathbbm{T} \ket{y_\pm^1(p) y_\pm^1(q)}=\ell_1^y \ket{y_\pm^1(p) y_\pm^1(q)}\ , \\  &&
\ell^z_1=-\frac{1}{g}l_1 + \frac{2}{g^2}\Theta_{\pm \pm}\ ,\qquad
\ell^y_1= \frac{1}{g}l_1 + \frac{2}{g^2}\Theta_{\pm \pm} \ , 
\eea
where the phases $\Theta_{++}=\Theta_{--}$ and $\Theta_{+-}=\Theta_{-+}$ are the two BOSST  \cite{Beccaria:2012kb,Borsato:2012ss,Borsato:2013hoa,Abbott:2013ixa} phases, see  (\ref{eq:BOSST-phase}). 

For scattering of bosonic particles with opposite U(1) charges  we find
\bea 
&&\mathbbm{T} \ket{z_\pm^1(p) z_\mp^1(q)} = \ell_2^{z} \ket{z_\pm^1(p) z_\mp^1(q)}+\dots,\qquad
\mathbbm{T} \ket{y_\pm^1(p) y_\mp^1(q)}=\ell_2^y \ket{y_\pm^1(p) y_\mp^1(q)}+\dots\ \ \nn
\\ &&
\ell_2^z= -\frac{1}{g}l_2 + \frac{2}{g^2}\Theta_{\pm \mp} , \qquad\ \ \ 
\ell_2^y= \frac{1}{g}l_2 + \frac{2}{g^2}\Theta_{\pm \mp}\ , 
\eea
which is again  finite after  the wave-function renormalization (\ref{eq:Zs}). Finally, for processes mixing the two bosonic coordinates we find\footnote{Note that in our conventions $y_+^1$ and $z_+^1$ have the same sign of the charge which differs from the convention used in
  \cite{Borsato:2013hoa}.}
\bea &
\mathbbm{T} \ket{z_\pm^1(p) y_\mp^1(q)} = \ell_3^{+-}\ket{z_\pm^1(p) y_\mp^1(q)}+\dots,\quad
\mathbbm{T} \ket{z_\pm^1(p) y_\pm^1(q)} = \ell_3^{++} \ket{z_\pm^1(p) y_\pm^1(q)}+\dots\nn
\\  &
\ell_3^{+-}=  -\frac{1}{g}l_3 + \frac{2}{g^2}\Theta_{\pm \pm}\ ,\qquad\ \ \ \
\ell_3^{++}= -\frac{1}{g}l_3 + \frac{2}{g^2}\Theta_{\pm \mp}\ . 
\eea
This amplitude is finite even  before using eq. (\ref{eq:Zs}), as expected from the fact that it 
mixes $z$ and $y$ particles.

\subsection{$AdS_2\times S^2\times T^6$}
The  difference compared to the $n=5,3$ cases 
 is that here  the massive bosons, which we parameterize with two real coordinates $x_1$ and $x_2$,
  are  neutral under the $U(1)$ symmetries  left after the light-cone gauge fixing.
  
For the  amplitudes  with bosonic in-states we find 
\bea \nn
&& \mathbbm{T}\ket{x^1(p)x^1(q)}=\ell_1^{x_1} \ket{x^1(p)x^1(q)}+\ell_3^{x_1} \ket{\chi_\pm^1(p) \chi_\mp^1(q)}+\dots,\\ \nn
&& \mathbbm{T}\ket{x^2(p)x^2(q)}=\ell_1^{x_2} \ket{x^2(p)x^2(q)}+\ell_3^{x_2} \ket{\chi_\pm^1(p) \chi_\mp^1(q)}+\dots, \\ \nn
&& \mathbbm{T}\ket{x^1(p) x^2(q)} = \ell_2 \ket{x^1(p) x^2(q)}+\ell_4\ket{\chi_\pm^1(p)\chi_\mp^1(q)}+\dots\ , 
\eea
where
\bea 
&& \ell_1^{x_1}=-\frac{1}{g}l_1 + \frac{4}{g^2}\Theta_{HL},\qquad\qquad 
\ell_1^{x_2}=\frac{1}{g}l_1 + \frac{4}{g^2}\Theta_{HL}, \\ 
&& \ell_3^{x_1}=-\frac{1}{g} l_3 ,\qquad
\ell_3^{x_2}=\frac{1}{g} l_3 ,
 \qquad \ell_2=  -\frac{1}{g} l_2 + \frac{4}{g^2}\Theta_{HL},\qquad \ell_4= -\frac{1}{g} l_4\  . 
\eea
Here we have  already implemented the wave-function renormalization 
 (\ref{eq:Zs}) (which, as was already mentioned earlier, 
  differs by a factor of 2  from the $n=5,3$ cases).
  
The   mixed  $BF \to BF$  amplitudes are   also finite after wave-function renormalization, 
\bea
&&\nn  \mathbbm{T}\ket{x^1(p) \chi_\pm(q)} = \ell_5^{x_1} \ket{x^1(p) \chi_\pm(q)}\ ,\qquad \ \ \ \mathbbm{T}\ket{x^2(p) \chi_\pm(q)} = \ell_5^{x_2} \ket{x^2(p) \chi_\pm(q)}\ , \\ &&
\ell_5^{x_1} = -\frac{1}{g} l_5\ ,\qquad\ \ \ \ 
\ell_5^{x_2} = \frac{1}{g} l_5 \ .
\eea

\section{Conclusions}
We have addressed the long standing question of how to properly compute 
the one-loop S-matrix of the 
$AdS_n \times S^n \times T^{10-2n}$ superstring around the BMN vacuum. By analyzing separately the one-loop  1-PI
 contribution to the two-particle scattering amplitude  and the off-shell one-loop two-point functions of massive fields 
 we demonstrated that the UV-divergences 
that appear should be interpreted as wave-function renormalization for the bosonic coordinates. 
Once this is taken into account 
the final expression for the one-loop S-matrix is  UV  finite.

We have  also outlined a regularization scheme which is consistent with the classical worldsheet 
symmetries.  One-loop computations in this scheme fully reproduce all known results about the massive  S-matrix 
predicted  by symmetries and integrability.
For the $n=2,3$ theories we found that the massless  loop  contributions to massive  two-particle 
 scattering amplitudes cancel out at one loop order. 
Thus, somewhat surprisingly, the massive 
sector S-matrix of the full superstring coincides with the one obtained from the  $AdS_n \times S^n $  supercoset sigma-model. 
Our result lends support to the generalized unitarity-based prescription of \cite{Bianchi:2014rfa} which also   leads to   a 
decoupling of massless modes at one loop for strings in $AdS_3 \times S^3 \times T^{4}$.

We  initiated a comparative study of  the light-cone and conformal gauge approaches to the one-loop S-matrix.
  While the former is 
well studied, the latter remains largely unexplored. A technical problem in conformal gauge is the presence of the unphysical 
massless longitudinal modes whose correct treatment remains to be understood. However, for the  $SU(2)$ sector  of the S-matrix 
we found evidence that accounting for the massless   modes  should be equivalent to passing from the BDS S-matrix (with no phase) to the 
S-matrix dressed with the standard AFS/HL/BES phase.

In conformal gauge  the one-loop two-point function for the bosons  happens to receive  a finite correction on-shell. 
This stands in contrast to the vanishing result in the light-cone gauge and suggests that the symmetries of the BMN vacuum 
have a different realization in the conformal gauge. For example, in conformal gauge the worldsheet energy is no longer related 
to the target space energy  and thus to the spin chain magnon dispersion relation of the dual gauge theory. 
The two-dimensional symmetries preserved by 
the BMN solution may lead to an extension of the non-local symmetries generated by the Lax connection and may ultimately 
determine the exact worldsheet  spectrum, perhaps along the lines of \cite{Belavin:1992en, Curtright:1992kp}.

One  interesting extension of our work is to  the two-loop order of the light-cone gauge-fixed  superstring 
around the BMN vacuum. A first step 
in this direction is the computation of the two-loop correction to the two-point function. Apart from  checking the strong 
coupling expansion of the magnon dispersion relation,  this   should  give a valuable insight into the  extension
of our regularization procedure to higher loops. It  should also  shed light on the issue of  (non)decoupling of massless 
modes at higher loops. Unitarity-based arguments suggest that the massless modes are no longer decoupled at three 
loops \cite{Engelund:2013fja} in the S-matrix. Two-loop dispersion relation calculations in the near flat space limit  
\cite{Murugan:2012mf} suggest   that massless modes may not decouple already at the two-loop level. 

It would also be very interesting to extend the analysis of this paper to  the $AdS_3 \times S^3 \times T^4$  superstring   with 
mixed NSNS and RR-flux \cite{Cagnazzo:2012se,Hoare:2013pma, Hoare:2013ida,Hoare:2013lja,Babichenko:2014yaa}. 
In \cite{Engelund:2013fja, Bianchi:2014rfa} the one-loop dressing phase for this theory 
was obtained via generalized unitarity methods. It 
would be very interesting to reproduce this result from an explicit worldsheet calculation and thus justify in the mixed flux case
the prescription for the treatment of the singular cuts.

\section*{Acknowledgments}
We thank S. Penati for useful discussions, and B. Hoare  and O. Ohlsson Sax  for useful discussions  and comments on the draft.
The work of R. Roiban is supported by the US Department of Energy under contract DE-SC0008745.
The work of P. Sundin was supported by a joint INFN and Milano-Bicocca postdoctoral grant.
The  work of  A. Tseytlin and L. Wulff  
was supported by the ERC Advanced grant No.290456
``Gauge theory -- string theory duality''
and also by the STFC grant
ST/J000353/1.

\

\appendix
\section{Reduction of one-loop integrals}\label{sec:int-reduction}
Using the identities (\ref{eq:bubbles-id1}) and (\ref{eq:bubbles-id2}) together with the assumption that we are allowed to shift the loop variable in $T^{rs}(P)$, we can rewrite the tadpole integrals as
\bea \nn
&& T^{12}(P)=P_- T^{11}(0)+ P_- P^2 T^{00}(0),\qquad
T^{21}(P)=P_+ T^{11}(0)+ P_+ P^2 T^{00}(0), \\ 
&& T^{11}(P)=T^{11}(0)+P^2 T^{00}(0),\qquad T^{01}(P)=P_- T^{00}(0),\qquad T^{10}(P)=P_+ T^{00}(0)\ .  \ \ \ \
\eea
For the  bubble integrals\footnote{Note   that for bubble-type integrals in the light-cone gauge, the 
two virtual particles always come with the same mass.}  
with only left- or right-moving momenta in the numerator  we get 
\bea
&& B^{03}(P)=\frac{P_-^2}{2P_+}\left(P^2-3m^2\right)B^{00}(P),\qquad
B^{30}(P)=\frac{P_+^2}{2P_-}\left(P^2-3m^2\right)B^{00}(P),\nn  \\ 
&& B^{02}(P)=-\frac{P_-}{P_+}\big(m^2-\frac{1}{2}P^2\big)B^{00}(P),\qquad
B^{20}(P)=-\frac{P_+}{P_-}\big(m^2-\frac{1}{2}P^2\big)B^{00}(P),\nn \\ 
&& B^{01}(P)=\frac{1}{2}P_- B^{00}(P),\qquad\qquad B^{10}(P)=\frac{1}{2}P_+ B^{00}(P)\ . 
\eea
Here we only recorded relations for the integrals that appear in the actual amplitudes (after using (\ref{eq:bubbles-id1})). 

\section{Expressions appearing in the light-cone gauge S-matrix}\label{sec:amplitudes}
Here we collect the explicit expressions for the amplitudes discussed  in section \ref{sec:scattering}.
 We will write some amplitudes in terms of   $\omega_p=\sqrt{ p^2 +1}$  and $p$, while others are written in terms of right-moving momenta $p_-=\omega_p-p$. 
  
\subsection{$AdS_5 \times S^5$}
For the tree-level amplitudes we have:
\bea 
\label{eq:ads5-ls} 
&& 
l_1=\frac{1}{2} \frac{(p+q)^2}{\omega_q p- \omega_p q}, \qquad l_2=\frac{1}{2}\big(\omega_q p + \omega_p q\big), \qquad 
l_3=-\frac{1}{4} \frac{(1-p_-^2)(1-q_-^2)}{\sqrt{p_- q_-}(p_-+q_-)}, \\ \nn 
&& l_4 = \frac{1}{4} \frac{(1-p_-q_-)(1-q_-^2)}{(p_--q_-)q_-}, \qquad 
l_5 = \frac{1}{4} \frac{(1+p_-q_-)(1-q_-^2)}{(p_-+q_-)q_-}, \qquad 
l_6 = \frac{1}{2} \frac{(1-p_-^2)(1-q_-^2)}{p_-^2-q_-^2}\ . 
\eea 
The one-loop Hernandez-Lopez phase term   in our notation  is 
\bea 
\label{eq:HL-phase}
\Theta_{HL}=\frac{1}{16\pi}\frac{(1-p_-^2)^2(1-q_-^2)^2(p_-^2+q_-^2)}{p_-q_-(p_-^2-q_-^2)^{2}}\log\frac{p_-}{q_-}
-\frac{1}{16\pi}\frac{(1-p_-^2)^2(1-q_-^2)^2}{p_-q_-(p_-^2-q_-^2)}.
\eea

\subsection{$AdS_3 \times S^3 \times T^4$}
The tree-level amplitudes  are 
\bea 
\label{eq:ads3-amplitudes}
l_1= \frac{1}{2}\frac{(p+q)^2}{\omega_q p - \omega_p q}, \qquad 
l_2=\frac{1}{2}\frac{(p-q)^2}{\omega_q p - \omega_p q}, \qquad 
l_3=\frac{1}{2}\big(\omega_p q + \omega_q p\big)\ .
\eea
The two one-loop phases, written in our notation, are 
\bea 
\label{eq:BOSST-phase} 
 \Theta_{\pm\pm}
 =\frac{1}{32\pi}\frac{(1-p_-^2)^2(1-q_-^2)^2}{p_-q_-(p_--q_-)^2}\log\frac{p_-}{q_-}-\frac{1}{64\pi}\frac{(p_-+q_-)(1-p_-^2)(1-q_-^2)(1-p_-q_-)^2}{p_-^2q_-^2(p_--q_-)}, \\ \nn 
 \Theta_{\pm\mp}=\frac{1}{32\pi}\frac{(1-p_-^2)^2(1-q_-^2)^2}{p_-q_-(p_-+q_-)^2}\log\frac{p_-}{q_-}+\frac{1}{64\pi}\frac{(p_--q_-)(1-p_-^2)(1-q_-^2)(1+p_-q_-)^2}{p_-^2q_-^2(p_-+q_-)}\ , 
\eea
which satisfy 
\bea \Theta_{\pm\pm}+\Theta_{\pm\mp}=\Theta_{HL} \ . \eea

For the expressions relevant to the $AdS_2 \times S^2 \times T^6$ case  we refer to \cite{Abbott:2013kka}.

\section{Comments on near BMN  S-matrix  in conformal gauge}

While the relation between worldsheet S matrix and 
gauge theory anomalous dimensions  of ``long" operators  described by the asymptotic Bethe ansatz 
 is best understood in a physical light-cone  type ``mixed" gauge adapted to the BMN vacuum
 (with $p^+$ or BMN charge being fixed in a uniform way) 
it is nevertheless interesting to explore   if  a similar relation may be 
formulated in the conformal gauge.  
There is a conceptual problem 
 in establishing such a relation stemming from the fact that, 
in conformal gauge, 
the worldsheet theory has two unphysical (longitudinal) massless modes. 
Their correct treatment (should  they be integrated out or should  they  be considered  as external states of the  S-matrix, etc.)  remains to be  understood. 
In this appendix we shall present results of some  computations 
that may help clarify these issues. 

Our tree-level S-matrix calculations below suggest that, at least 
in the $SU(2)$ sector, the  correct treatment of the massless modes 
should be equivalent to passing from the  (strong coupling limit of the) 
``phaseless" BDS  \cite{Beisert:2004hm} 
  S-matrix to the S-matrix dressed with the AFS/BES  \cite{Arutyunov:2004vx,Beisert:2006ez} phase.

\def \td {\tilde}

\subsection{Tree-level bosonic  S-matrix in the $AdS_n\times S^n\times T^{10-2n}$ theory \label{CG_bose_S_matrix}}
Let us start with fixing  the conformal gauge in the string action \rf{ac}
\begin{equation}
\sqrt{-h} h^{ij} =\eta^{ij}
\end{equation}
and then expand the  Lagrangian around the BMN solution
 $x^+\equiv \frac12(t+\varphi)=\tau\ $ with  \ $x^-,z^m,y^m=0$.\footnote{As is well known, the
 $x^+=\tau$ condition cannot be viewed as an analog of flat-space l.c.  gauge that fixes remaining 
 conformal reparametrizations as it does not solve the string equations  for generic ``transverse" 
 string coordinates of $AdS_n \times S^n$ space.}
 Setting $t=\tau + \td t, \  \varphi=\tau + \td \varphi$  and expanding 
  to quartic order,  the bosonic
Lagrangian  in the coordinates \rf{eq:AdSmetric},\rf{eq:Smetric}  becomes  the sum of three terms 
\bea
{\cal L}_{B2} &=&\frac{1}{2}(\, - \partial_i \td t \partial^i \td t    + \partial_i \td \varphi \partial^i \td \varphi\, ) 
  +    \frac{1}{2}(\, \partial_i z_m\partial^i z^m + \partial_i y_m\partial^i y^m -  z^2 - y^2\,)
\\
{\cal L}_{B3} &=&- \partial_0 \td t \, z^m z_m -  \partial_0\td  \varphi\,  y^m y_m 
\\
{\cal L}_{B4} &=& -\frac{1}{2}  \partial_i \td t\partial^i \td t\, z^2 - \frac{1}{2}\partial_i\td \varphi \partial^i\td\varphi \,y^2
              + \frac{1}{4} \big[ z^2\,\partial_i z_m\partial^i z^m  -  \partial_i y_m \partial^i y^m \, y^2
             - (z^2)^2 + (y^2)^2\big]\ \ \ 
\label{LB234}             
\eea
where  $z^2 \equiv z^m z_m$, etc.,   and the index  $m$ runs over the transverse directions. We 
choose the fields to be real to interpolate easily between theories   with different dimensions
of $AdS \times S$.

Due to  the presence of cubic interaction terms involving one massless longitudinal field,
 the $t$-channel 
contribution to the S matrix is singular on shell. We regularize this singularity as follows: 
\begin{enumerate}
\item introduce a small regulating mass (as for one-loop IR-divergent integrals)
\item compute the off-shell four-point Green's function 
\item put the Green's function on shell and amputate
\item take the regulating mass to zero
\end{enumerate}
\noindent 
The result of  this prescription is a finite tree-level S-matrix. 

The  ``transverse" $SO(n-1)\times SO(n-1)$ symmetry of the Lagrangian as well as the decoupling of the  $AdS_n $ and $S^n$  fluctuations in the conformal gauge require 
that the S-matrix takes the general form
\begin{align}
\mathbbm{T}|z^m(p)z^n(q)\rangle &= ~~ (A \delta_k^m\delta_l^n + B \delta_k^n\delta_l^m+C\delta^{mn}\delta_{kl})|z^k(p)z^l(q)\rangle
\cr
\mathbbm{T}|y^m(p)y^n(q)\rangle  &= -(A \delta_k^m\delta_l^n + B \delta_k^n\delta_l^m+C\delta^{mn}\delta_{kl})|y^k(p)y^l(q)\rangle
\cr
\mathbbm{T} |y^m(p)z^n(q)\rangle  &= 0
\label{TTmatrix}
\ .
\end{align}
Using the above prescription for the  massless modes and a relativistic normalization for the  S-matrix 
the free coefficients in \rf{TTmatrix} are given by
\begin{equation}
A=0\ , 
\qquad\qquad 
B =-C = 4 pq \ .
\end{equation}
With a non-relativistic normalization and with a manifestly solved momentum-conservation  
constraint the  above coefficients become
\begin{equation}
A=0\ , 
\qquad\qquad 
B =-C = \frac{4 pq}{p\omega_{q} - q\omega_{p}  } \ .
\end{equation}
This S-matrix is, of course, consistent with the classical Yang-Baxter equation.

In the case of $AdS_5\times S^5$,  integrability together with the fact that $SO(4)\simeq SU(2)\otimes SU(2)$ 
require that 
\begin{equation}
\mathbbm{T} = \unit\otimes {\rm T} + {\rm T}\otimes 1\ , 
\label{factorized}
\end{equation}
where $\unit$ and ${\rm T}$ act on $SU(2)$ indices from the decomposition of the two $SO(4)$ factors as
\begin{equation}
\label{}
\unit_{ab}^{cd}=\delta _b^d\delta _a^c\ , \qquad
\qquad
P_{ab}^{cd}=\delta _b^c\delta _a^d \ . 
\end{equation}
It is not difficult to see that the non-zero entries of $\mathbbm{T}$ may be written as
\begin{align}
\mathbbm{T} |z(p)z(q)\rangle  &= -\big[B \unit\otimes\unit - B(\unit\otimes P+P\otimes \unit)\big] \, |z(p)z(q)\rangle
\cr
\mathbbm{T} |y(p)y(q)\rangle &= ~\;~\big[B \unit\otimes\unit - B(\unit\otimes P+P\otimes \unit) \big]\,  |y(p)y(q)\rangle \ .
\end{align}
This is indeed  consistent with the factorized structure \eqref{factorized}.

Longitudinal states appear to scatter trivially off the massive states. This may be understood in two steps. First, the cubic terms 
may be eliminated by a non-local field redefinition. While potentially worrisome, the effect 
of the non-locality is only to generate effective quartic interaction terms between massive fields which correspond to the 
Feynman graphs with exchange of longitudinal fields. The second step is to notice that momentum conservation implies 
that the S-matrix elements following from the quartic terms are proportional to the dispersion relation for the longitudinal 
fields and thus vanish on shell. Such trivial scattering of  longitudinal modes
may not be unexpected   given that  for massless 
fields  it is  notoriously difficult to define a consistent scattering
 theory that has a perturbative regime.

Clearly, the S-matrix \eqref{TTmatrix} is different from the one obtained in 
the ``light-cone''  $a$-gauge \cite{Klose:2006zd}. While the latter has  
nontrivial $yz\rightarrow yz$ matrix elements, the former does not. Such matrix elements may be generated at loop level 
through fermion loops as well as loops of longitudinal modes. The non-zero matrix elements are also different; while the 
difference is proportional to the identity operator, $\unit \otimes \unit$, it is not only an overall phase as it affects differently 
the scattering of $AdS$ and $S$ fluctuations:
\begin{align}
\delta \mathbbm{T}_{z(p)z(q)\rightarrow z(p)z(q)} &= \frac{1}{2}\left[
(1-2a)(p\omega_{q} - q\omega_{p})
+\frac{p^2+q^2}{p\omega_{q} - q\omega_{p}}\right] \unit\otimes\unit 
\cr
\delta \mathbbm{T}_{y(p)y(q)\rightarrow y(p)y(q)} &= \frac{1}{2}\left[
(1-2a)(p\omega_{q} - q\omega_{p})
-\frac{p^2+q^2}{p\omega_{q} - q\omega_{p}}\right] \unit\otimes\unit \ .
\end{align}
Through generalized unitarity tree-level differences imply  \cite{Engelund:2013fja} that the one-loop S matrix in 
conformal gauge is also different from the one-loop S matrix in the $a$-gauge. 

It is interesting to note that,  when restricted to the $SU(2)$ sector, the S-matrix \eqref{TTmatrix} is the same as the BDS 
S-matrix in  the small momentum limit. It was suggested  in \cite{Rej:2007vm, Sakai:2007rk} that the dressing 
phase may be understood as a consequence of a nontrivial vacuum in the Bethe equations based on the BDS S-matrix.  This  may be viewed as  a hint  that the  difference 
  between the conformal gauge S-matrix and the  light-cone gauge S-matrix 
from the perspective of the usual asymptotic Bethe ansatz may be  due to  a nontrivial choice of 
vacuum for the longitudinal excitations once a consistent scattering theory is defined for the latter. 
A somewhat similar suggestion
was made for the non-transverse excitations of a  principal chiral model 
 on $\mathbb{R}\times S^3$ \cite{Gromov:2006cq} and of some  conformal 
sigma models \cite{Mann:2005ab}. In our case this interpretation is 
also supported  by the fact that in the presence of the 
longitudinal fields  the massive fields are potentially unstable, loosing energy by emitting low energy massless quanta.

\subsection{One-loop bosonic dispersion relation \label{app:conformal2pf}}
Apart from the S-matrix, 
the other essential ingredient of a Bethe ansatz  is the exact dispersion relation
 for the elementary 
excitations.  To one-loop order the quantum corrections to dispersion relation 
vanish in the $a=1/2$ gauge. As discussed in the main text,  computing  the 
off-shell two-point functions leads  to  a  nontrivial 
wave-function renormalization. It is interesting to carry out a similar study in the conformal gauge.

Let  us compute  the one-loop two-point function  by directly expanding 
around the BMN vacuum.  We 
will describe the calculation for the $AdS_5 \times S^5$  case  
and then comment on extension to lower-dimensional   cases.
Apart from the bosonic action to quartic order given  in Appendix~\ref{CG_bose_S_matrix}, we also need terms bilinear 
in fermions   and up to  quadratic order in bosons. 
They are obtained from the  $AdS_5 \times S^5$ action  in section 
\ref{GSactions}
by imposing the $\kappa$-symmetry  light-cone  gauge  $\Gamma^+\Theta=0$. 

There are in principle 
four graphs contributing to the two-point function of massive bosons: a bosonic bubble and a tadpole and also a 
fermionic bubble (which in our case vanishes identically) and a tadpole. The bosonic and fermionic contributions 
are separately divergent off shell, but the 
divergence is proportional to the classical  equation of motion  $(p_+p_--1)$ so they are finite on shell. There is a finite momentum dependent contribution to the two-point function which arises entirely from the the bosonic bubble graph:
\begin{equation}
i\Pi^{(1)} = 2\varepsilon \int\frac{d^2 q}{(2\pi)^2}\frac{q_+^2+q_-^2}{q^2((q+p)^2-1)}
\end{equation}
where $\varepsilon=\pm 1$ for the transverse AdS and sphere  fluctuations, respectively. These integrals, while logarithmically divergent 
by power counting, are finite in dimensional regularization. Evaluating them leads to\footnote{Note that there is  a nonzero imaginary part  
related to the presence of massless  states.}
\begin{align}
\Pi^{(1)}(p_+,p_-)&=~~\frac{\varepsilon}{4\pi}\frac{p_0^2+p_1^2}{p_+p_-}
\left(1- \frac{1-p_+p_- }{p_+p_-}\ln\frac{1-p_+p_- }{p_+p_-}\right)  \ .
\end{align}
 These expressions are non-vanishing on shell and they lead to a correction to the tree-level dispersion relation:
\begin{equation}
\omega^2 = 1+p^2+\frac{\varepsilon}{4\pi g}(1+2p^2) \ .
\end{equation}
The meaning of this correction and its effect on the symmetries of the S-matrix remain  to be clarified.  

To extend the above  $AdS_5\times S^5$ calculation  to other   
$AdS_n\times S^n\times T^{10-2n}$ cases we notice 
that the only non-vanishing contribution comes from the bosonic bubble
graph  whose internal-line field content is uniquely fixed by the 
choice of the external field. Thus, we conclude that the same two-point function should appear in all other cases.

\newpage 

\bibliographystyle{JHEP}
\bibliography{ads2ads3etc}
\end{document}